\documentclass[journal,article,submit,pdftex,moreauthors]{mdpi} 
\renewcommand{\linenumbers}{}

\usepackage{amssymb}
\usepackage{amsfonts}
\usepackage{amsmath}
\usepackage{amsthm}

\DeclareMathOperator{\Tr}{Tr}

\firstpage{1} 
\pubvolume{1}
\issuenum{1}
\articlenumber{0}
\pubyear{2025}
\copyrightyear{2025}
\datereceived{ } 
\daterevised{ } 
\dateaccepted{ } 
\datepublished{ } 
\hreflink{https://doi.org/} 

\Title{$D_N$-Orthogonal Freedom in the Canonical Seesaw: Flavor Invariants and Physical Non-Equivalence of F-Classes}

\TitleCitation{Title}

\Author{Jianlong Lu $^{1}$\orcidA{}}

\AuthorNames{Jianlong Lu}

\isAPAStyle{%
       \AuthorCitation{Lu, J.}
         }{%
        \isChicagoStyle{%
        \AuthorCitation{Jianlong Lu.}
        }{
        \AuthorCitation{Lu, J.}
        }
}

\address{%
$^{1}$ \quad Department of Mathematics, Faculty of Science, National University of Singapore}

\corres{Correspondence: jianlong@nus.edu.sg}

\abstract{We study basis-independent structures in the Type-I seesaw mechanism for
light Majorana neutrinos, assuming the canonical scenario with three heavy right-handed
(sterile) neutrinos. Let $m_\nu$ denote the $3\times3$ mass matrix of light neutrinos,
obtained at tree level from heavy Majorana singlets with diagonal mass matrix
$D_N = \mathrm{diag}(M_1,M_2,M_3)$ and Dirac matrix $m_D$. We show that all
right-actions $F$ on the seesaw matrix that leave $m_\nu$ unchanged form the group
$G = D_N^{1/2} O(3,\mathbb{C}) D_N^{-1/2}$. While oscillation data determine the PMNS
matrix $U_{\rm PMNS}$ and the mass-squared splittings, they do not fix the $F$-class
within $G$. We classify basis-invariant quantities into those that are class-blind
(e.g.\ $\det\eta$) and class-sensitive (e.g.\ $\mathrm{Tr}\,\eta$, $\mathrm{Tr}\,\eta^2$,
an alignment measure, and CP-odd traces relevant to leptogenesis), where $\eta$ denotes
the non-unitarity matrix of the light sector. We provide explicit formulas and both
high-scale and GeV-scale benchmark examples that illustrate these invariant fingerprints
and their scaling with $D_N$. This converts the degeneracy at fixed $m_\nu$ into
measurable, basis-invariant fingerprints.}


\begin{document}


\section{Introduction}
Neutrino oscillation experiments have established that the three active neutrinos are massive and mix, implying a Majorana mass matrix $m_\nu$ for the light states. Among the simplest and most predictive ultraviolet (UV) completions that generate $m_\nu$ is the canonical seesaw mechanism~\citep{Minkowski(1977),Gell-Mann(1979),Yanagida(1979),Glashow(1980),Mohapatra(1980)}, in which three heavy right-handed (sterile) neutrinos couple to the lepton doublets via Dirac Yukawas and carry large Majorana masses. In a basis with diagonal charged leptons, 

We define $m_\nu$ as the tree-level effective light-neutrino mass matrix obtained by integrating out the heavy fields, i.e. the Schur-complement (Weinberg operator) expression
\begin{equation}
m_\nu \equiv -\,m_D\,D_N^{-1}\,m_D^{T}.
\label{eq:mnu}
\end{equation}
This is the tree-level matching relation for the Weinberg operator, obtained by integrating out the heavy fields at order $1/D_N$. In what follows, ‘exact’ refers to algebraic identities within this tree-level seesaw framework, without further series expansions in small mixing angles or in the entries of $m_D$.
For the full $6 \times 6$ neutral-fermion mass matrix $\mathcal{M}\thickspace =\thickspace 
\begin{pmatrix}
0 & m_D\\[2pt]
m_D^{T} & M_R
\end{pmatrix}$, the Takagi decomposition implies the exact identity
\begin{equation}
m_\nu \thickspace =\thickspace  N\thinspace D_\nu\thinspace N^{T} \thickspace =\thickspace  -\thinspace R\thinspace D_N\thinspace R^{T},
\label{eq:intro-exact}
\end{equation}
where $N$ denotes the light--light mixing sub-block and $D_\nu=\mathrm{diag}(m_1,m_2,m_3)>0$ are the light eigenmasses, $R$ denotes the heavy--light mixing sub-block and $D_N=\mathrm{diag}(M_1,M_2,M_3)>0$ are the heavy eigenmasses. Eq.~\eqref{eq:intro-exact} emphasizes that low-energy information is encoded in the combination $R D_N R^{T}$ rather than in $R$ or $D_N$ separately. Eqs. ~\eqref{eq:mnu}–\eqref{eq:intro-exact} should be understood as tree-level relations in the usual seesaw sense. The effective mass matrix $m_\nu$ is defined by the Schur complement of the heavy fields and coincides with the light–sector combination $N D_\nu N^T$ in the full Takagi diagonalization. We neglect higher-dimensional operators and loop corrections throughout. Within this tree-level framework our use of ‘exact’ simply means that no further expansion in $m_D D_N^{-1}$ is performed.

Oscillation data determine the mixing angles, $\delta_{\rm CP}$, and the mass-squared splittings $(\Delta m^2_{21},|\Delta m^2_{31}|)$, thereby constraining $U_{\rm PMNS}$ and two independent mass–squared splittings of the eigenvalues of $D_\nu=\mathrm{diag}(m_1,m_2,m_3)$ \cite{PDG(2024),Esteban(2024)}. The absolute mass scale $m_{\rm lightest}$ and the two Majorana phases remain unconstrained by oscillations.
However, the UV completion in the seesaw is not uniquely determined by $m_\nu$: the map $(m_D,M_R)\mapsto m_\nu$ is many-to-one. This degeneracy appears transparently in Eq.~\eqref{eq:intro-exact}: if $F$ is any matrix obeying
\begin{equation}
F\thinspace D_N\thinspace F^{T} \thickspace =\thickspace  D_N,
\label{eq:intro-invariance}
\end{equation}
then $R\to R F$ leaves $m_\nu$ identical. Consequently, distinct pairs $(R,D_N)$ (or, equivalently, distinct $(m_D,M_R)$) can share the same $m_\nu$ and thus the same oscillation phenomenology, while differing in heavy--light observables (non-unitarity) \citep{Forero(2021),Hu(2021)}, charged-lepton flavor violation (LFV) \citep{Lindner(2018),Kuno(2001),Calibbi(2018)}, and leptogenesis \citep{Davidson(2008),Nardi(2006),Blanchet(2007)}.

A minimal and widely studied origin of $m_\nu$ is the type-I seesaw \cite{Minkowski(1977),Yanagida(1979),Gell-Mann(1979),Glashow(1980),Mohapatra(1980)} (see \cite{Smirnov(2006),King(2013)} for reviews). Structures of $m_\nu$ and $U_{\rm PMNS}$ have been explored via texture zeros and flavour symmetries \cite{Frampton(2002),Grimus(2001),Rodejohann(2005),Feruglio(2015),Harrison(2002)}, and through basis-invariant diagnostics of CP and alignment \cite{Jarlskog(1985),Rebelo(1986)}. Phenomenologically, the relevant probes include precision tests of non-unitarity \cite{Antusch(2006),Yasuda(2007),Blennow(2017),Aloni(2024)}, searches for charged-lepton flavour violation, and heavy-neutral-lepton (HNL) searches across energy scales \cite{Atre(2009),Drewes(2013),Deppisch(2015),Bolton(2025)}, as well as leptogenesis \cite{Ibarra(2002),Davidson(2008),Pilaftsis(2004),Abada(2006),Nardi(2006)}.

The experimentally favoured modulus relation $|U_{\mu i}| = |U_{\tau i}|$ $(i=1,2,3)$ has
motivated attempts to uncover an underlying flavour symmetry. In
particular, Xing~\cite{Xing(2022)} worked in the canonical seesaw
framework and combined the exact seesaw formula and the unitarity of the
full $6\times 6$ mixing matrix to argue that $|U_{\mu i}| = |U_{\tau i}|$ can imply
$|R_{\mu i}| = |R_{\tau i}|$. In the special scenario $U = P\,U^{*}$ with
$P$ the $\mu\leftrightarrow\tau$ permutation, he further obtained the stronger
condition
\begin{equation}
  R' = P_{23} R^{\prime *}\,,
\end{equation}
which realizes a (generalized) $\mu\text{–}\tau$ reflection symmetry at the seesaw
scale and leads to constrained textures for $M_D$ and $M_R$. In this
“minimal $\mu\text{–}\tau$” picture, the seesaw flavour structure is essentially
fixed once low–energy data and $\mu\text{–}\tau$ reflection are assumed.\\
In our previous work~\cite{Jianlong(2024)} we revisited the same
starting point and asked whether the implication
$|U_{\mu i}| = |U_{\tau i}| \Rightarrow R = P R^{*}$ is in fact unique within the
canonical seesaw. Using the exact identity
$m_\nu = - R D_N R^{T}$ together with the right–action freedom
\begin{equation}
\label{eqc}
  F D_N F^{T} = D_N \quad \Longrightarrow \quad
  R \;\to\; R F
\end{equation}
we constructed explicit, non–trivial matrices $F$ for which
$(R F) D_N (R F)^{T} = R D_N R^{T}$ reproduces the same $m_\nu$ and thus the
same $|U_{\mu i}| = |U_{\tau i}|$, while $R F$ does \emph{not} satisfy
$R F = P (R F)^{*}$. This showed that the minimal $\mu\text{–}\tau$ realization
studied in Ref.~\cite{Xing(2022)} is only one particular point in a larger space of
seesaw completions compatible with the same low–energy data.\\
The present work promotes that observation to a systematic, basis–invariant
classification of the $m_\nu$-preserving freedom.

Our first main result is a group-theoretic classification of all $F$ that satisfy Eq.~\eqref{eq:intro-invariance}:
\begin{equation}
\mathcal{G} \thickspace \equiv\thickspace  \{F\in GL(3,\mathbb{C})\mid F D_N F^{T}=D_N\}
\thickspace =\thickspace  D_N^{1/2}\thinspace O(3,\mathbb{C})\thinspace D_N^{-1/2},
\label{eq:intro-classification}
\end{equation}
obtained by setting $H=D_N^{-1/2} F D_N^{1/2}$ so that $H H^{T}=I$. Thus, the entire $m_\nu$-preserving freedom is the conjugate of the complex orthogonal group by $D_N^{1/2}$. This perspective elevates the previously constructed examples to representatives inside a continuous symmetry manifold and clarifies how the freedom enlarges when heavy neutrino masses are degenerate.

Our second main result is to separate \emph{class-blind} from \emph{class-sensitive} observables using weak-basis (flavor) invariants. 
We call a weak-basis invariant class-blind if it is unchanged under $G$ at fixed $m_\nu$ (e.g., $\det\eta$), and class-sensitive otherwise (e.g., $\mathrm{Tr},\eta$). Unlike our Refs.\cite{Jianlong(2024),Jianlong(2025)}, which catalogued allowed textures, we provide a group-theoretic classification of the right-action freedom and a phenomenological partition of invariants by their transformation under $G$. Any scalar built solely from $h_\nu\equiv m_\nu m_\nu^\dagger$ is invariant under weak-basis transformations (WBT) and is unchanged by $F\in\mathcal{G}$. Examples include $\Tr h_\nu$, $\Tr h_\nu^2$, and $\det h_\nu$ (as well as the low-energy Jarlskog invariant). These provide internal consistency checks across classes. By contrast, the non-unitarity matrix $\eta\equiv\tfrac12 R R^\dagger$ and Dirac-sector combinations distinguish classes. We construct WBT-invariant scalars that change with $F$ in general: $\Tr\eta$, $\Tr\eta^2$, the alignment invariant $\Tr([\eta,h_\nu]^2)$, and a CP-odd leptogenesis invariant $\mathcal{I}_{\rm CP}^{(1)}$ built from $H_D\equiv m_D^\dagger m_D$ and $M_R$. A notable exception is $\det\eta = 2^{-3}\det(D_\nu)\det(D_N^{-1})$, which is class-independent. We provide an analytic proof.

Because all classes in $\mathcal{G}$ reproduce the same $m_\nu$, oscillation experiments alone cannot select a unique completion. The $m_\nu$-preserving families in $\mathcal{G}$ are distinguished by the basis-invariant diagnostics $\eta\equiv\tfrac12 RR^\dagger$ and its flavor pattern, $H_D\equiv m_D^\dagger m_D$, and CP-odd traces such as $\mathcal{I}_{\rm CP}^{(1)}$ with flavored extensions.
These map directly onto observables: precision tests of non-unitarity bound $\eta$, searches for charged–lepton flavor violation probe entries of $H_D$, and leptogenesis is controlled by $\mathcal{I}_{\rm CP}^{(1)}$. Taken together, these measurements constrain, and can ultimately select, the viable UV completions within $\mathcal{G}$ that otherwise predict the same $m_\nu$.
We illustrate these connections with a set of representative benchmarks from these families: keeping $(U_{\rm PMNS},D_\nu,D_N)$ fixed, the class-sensitive invariants separate by factors of a few for $\Tr\eta$ and $\Tr\eta^2$ up to many orders of magnitude for the alignment invariant, and $\mathcal{I}_{\rm CP}^{(1)}$ may flip sign across classes, while the class-blind set remains identical. We also explain how heavy-mass degeneracies enlarge $\mathcal{G}$ and suppress unflavored CP-odd invariants, motivating flavored generalizations.

Ref.\cite{Xing(2022)} effectively fixed one representative within the different F-classes by imposing $|R_{\mu i}|=|R_{\tau i}|$ (equivalently, $R=P R$ in our notation), thereby discarding the other classes. Our classification clarifies that this is a consistent choice of class, not a general identity.

The rest of the paper is organized as follows. Section~\ref{sec:setup} reviews the exact tree-level matching and notation. Section~\ref{sec:classification} proves the classification in Eq.~\eqref{eq:intro-classification}. Section~\ref{sec:invariants} develops the invariant machinery, proving flavor invariance and identifying class-blind versus class-sensitive structures, including the class-independence of $\det\eta$. Section~\ref{sec:benchmark} presents a benchmark study based on six representative points
$A$--$F$ and discusses the resulting numerical patterns. Section~\ref{sec:degeneracy} analyzes heavy-mass degeneracies and flavor invariants. Phenomenological implications are discussed in Section~\ref{sec:pheno}, and conclusions and outlook appear in Section~\ref{sec:conclusions}.

\section{Seesaw setup and diagonalization}
\label{sec:setup}
We work in the charged-lepton mass basis. The canonical seesaw Lagrangian contains
\begin{equation}
\mathcal{L}\supset -\thinspace \overline{\ell_L}\thinspace m_D\thinspace N_R \thickspace -\thickspace  \tfrac12\thinspace N_R^{T} C\thinspace M_R\thinspace N_R \thickspace +\thickspace  \text{h.c.},
\end{equation}
with a complex $3\times 3$ Dirac mass matrix $m_D$ and a complex symmetric Majorana mass matrix $M_R$. It is convenient to assemble these into the exact $6\times 6$ neutral-fermion mass matrix
\begin{equation}
\mathcal{M}\thickspace =\thickspace 
\begin{pmatrix}
0 & m_D\\[2pt]
m_D^{T} & M_R
\end{pmatrix}.
\end{equation}
At tree level, one can perform a standard (in general non-unitary) block
diagonalization of the $6\times 6$ mass matrix $\mathcal{M}$. There exists a matrix of the form
\begin{equation}
  W =
  \begin{pmatrix}
    \mathbf{1} & \Theta \\
    0          & \mathbf{1}
  \end{pmatrix},
  \qquad \Theta = \mathcal{O}(m_D M_R^{-1}),
\end{equation}
such that
\begin{equation}
  W^T \mathcal{M} W =
  \begin{pmatrix}
    m_\nu & 0 \\
    0     & M_R + \mathcal{O}(m_D^2 M_R^{-1})
  \end{pmatrix},
  \qquad m_\nu = -\,m_D D_N^{-1} m_D^{T}.
\end{equation}
This step defines the tree-level effective mass matrix $m_\nu$ in Eq.~\ref{eq:mnu} as
the Schur complement of the heavy block $M_R$. The matrix $W$ is in general not unitary
and therefore does \emph{not} correspond to an exact change of basis in the
full theory. The exact unitary diagonalization is instead provided by the
matrix $\mathcal{U}$ in Eq.~\ref{eq:U}.\\
Since $M$ is complex symmetric, it admits a Takagi-type factorization. There exists a unitary matrix
\begin{equation}
\mathcal{U}\thickspace =\thickspace 
\begin{pmatrix}
N & R\\[2pt]
S & T
\end{pmatrix},
\qquad
\mathcal{U}^{\dagger}\mathcal{U}=\mathbb{I}=\mathcal{U}\mathcal{U}^{\dagger},
\end{equation}
such that
\begin{equation}
\mathcal{U}^{\dagger}\thinspace \mathcal{M}\thinspace \mathcal{U}^{*}
\thickspace =\thickspace 
\mathrm{diag}(D_\nu, D_N),
\qquad
D_\nu=\mathrm{diag}(m_1,m_2,m_3)\ge 0,\quad
D_N=\mathrm{diag}(M_1,M_2,M_3)>0.
\label{eq:U}
\end{equation}
This is equivalent to the usual Takagi factorization
$\mathcal{V}^T \mathcal{M} \mathcal{V} = \mathrm{diag}(D_\nu,D_N)$, with $\mathcal{V} = \mathcal{U}^*$.

Block unitarity of $\mathcal{U}$ implies the identities
\begin{equation}
N N^\dagger + R R^\dagger = \mathbb{I},\qquad
S S^\dagger + T T^\dagger = \mathbb{I},\qquad
N S^\dagger + R T^\dagger = 0,
\label{eq:block-unitarity}
\end{equation}
and their Hermitian conjugates. The $3\times 3$ sub-block $R$ encodes heavy–light mixing, with the standard non-unitarity measure \cite{Blennow(2017)}
\begin{equation}
\eta \thickspace \equiv\thickspace  \tfrac12\thinspace R R^\dagger .
\end{equation}
Projecting the diagonalization condition $\mathcal{U}^{\dagger}\thinspace \mathcal{M}\thinspace \mathcal{U}^{*}=\mathrm{diag}(D_\nu,D_N)$ onto the active–active block yields the exact identities \cite{Jianlong(2024)}
\begin{equation}
\label{eq:exact-id}
m_\nu \thickspace =\thickspace  N\thinspace D_\nu\thinspace N^{T} \thickspace =\thickspace  -\thinspace R\thinspace D_N\thinspace R^{T},
\end{equation}
which make explicit that low-energy information is encoded in the combination $R D_N R^{T}$ rather than in $R$ or $D_N$ separately. Eq.~\eqref{eq:exact-id} is an exact identity following from the Takagi relation in Eq.~\ref{eq:U}. It is obtained without performing any further series expansion in $m_D/M_R$ beyond the tree-level mass matrix $\mathcal{M}$ itself. The Takagi factorization $\mathcal{U}^{\dagger}\thinspace \mathcal{M}\thinspace \mathcal{U}^{*}=\mathrm{diag}(D_\nu,D_N)$ then provides the exact unitary change of basis from flavor to mass eigenstates. Eq.~\ref{eq:intro-exact} follows by projecting onto the light sector. $U_{\rm PMNS}$ is obtained from $N$ up to Majorana phases in the charged-lepton mass basis. Oscillations fix $(\theta_{12},\theta_{13},\theta_{23},\delta_{\rm CP})$ and $(\Delta m^2_{21},|\Delta m^2_{31}|)$, while $m_{\rm lightest}$ and Majorana phases remain free. For normal (inverted) ordering, $(m_1,m_2,m_3)$ follow from $m_{\rm lightest}$ and the measured splittings. 

For later use it is convenient to introduce the Hermitian combinations
\begin{equation}
h_\nu \thickspace \equiv\thickspace  m_\nu m_\nu^\dagger,\qquad
H_D \thickspace \equiv\thickspace  m_D^\dagger m_D,\qquad
X \thickspace \equiv\thickspace  M_R^\dagger M_R,
\end{equation}
which appear in weak-basis (flavor) invariants. Under weak-basis transformations that preserve canonical kinetic terms,
\begin{equation}
\ell_L \to W_L\thinspace \ell_L,\qquad e_R \to W_e\thinspace e_R,\qquad N_R \to W_R\thinspace N_R,
\qquad W_{L,e,R}\in U(3),
\end{equation}
the mass matrices transform as
\begin{equation}
m_D \to W_L\thinspace m_D\thinspace W_R^\dagger,\qquad
M_R \to W_R^{*}\thinspace M_R\thinspace W_R^{\dagger},
\label{eq:WBT}
\end{equation}
implying
\begin{equation}
m_\nu \to W_L\thinspace m_\nu\thinspace W_L^{T},\qquad
h_\nu \to W_L\thinspace h_\nu\thinspace W_L^\dagger,\qquad
H_D \to W_R\thinspace H_D\thinspace W_R^\dagger,\qquad
X \to W_R\thinspace X\thinspace W_R^\dagger .
\end{equation}
With the block unitary $V\equiv \mathrm{diag}(W_L,\,W_R^{*})$, the mass matrix transforms by congruence
$\mathcal M\to\mathcal M' \equiv V^{T}\mathcal M V$.
A diagonalizer of $\mathcal M'$ is then $\mathcal{U}' \equiv V^\dagger \mathcal{U}\,\mathcal P$ with
$\mathcal P\equiv \mathrm{diag}(P_\nu,P_N)$ (diagonal phase/permutation matrices used to keep $D_\nu,D_N$ real–positive and ordered).
Therefore the Takagi blocks transform as
\begin{align}
N' &= W_L^\dagger\,N\,P_\nu,\\
R' &= W_L^\dagger\,R\,P_N,\\
S' &= W_R^{T}\,S\,P_\nu,\\
T' &= W_R^{T}\,T\,P_N \;.
\end{align}

In particular, any scalar built solely from $h_\nu$ (such as $\Tr h_\nu$, $\Tr h_\nu^2$, $\det h_\nu$) is invariant under weak-basis transformations. Similar remarks apply to traces built from $H_D$ and $X$ that are arranged to be basis invariant.

It is often useful to parameterize $m_D$ in terms of low-energy data and a complex orthogonal matrix. When $M_R$ is diagonal and positive, one may write the Casas–Ibarra form \cite{Casas(2001)}
\begin{equation}
m_D \thickspace =\thickspace  i\thinspace N\thinspace \sqrt{D_\nu}\thinspace \Omega\thinspace \sqrt{D_N},
\qquad \Omega \in O(3,\mathbb{C}),\qquad \Omega\thinspace \Omega^{T}=\mathbb{I},
\label{eq:CI}
\end{equation}
which automatically reproduces $m_\nu =  N \sqrt{D_\nu}\thinspace \Omega\thinspace \Omega^{T}\sqrt{D_\nu}\thinspace N^{T} =  N D_\nu N^{T}$. In terms of the block $R$ one obtains
\begin{equation}
R \thickspace =\thickspace  i\thinspace N\thinspace \sqrt{D_\nu}\thinspace \Omega\thinspace D_N^{-1/2},
\label{eq:R-omega}
\end{equation}
and the identity $m_\nu=-R D_N R^{T}$ follows immediately. Right multiplication of $R$ by a matrix $F$ that satisfies
\begin{equation}
F\thinspace D_N\thinspace F^{T} = D_N
\label{eq:FDNF}
\end{equation}
is equivalent, within the parametrization Eq.~\eqref{eq:R-omega}, to the replacement $\Omega\to \Omega H$ with
\begin{equation}
H \thickspace \equiv\thickspace  D_N^{-1/2}\thinspace F\thinspace D_N^{1/2}\ \in\ O(3,\mathbb{C}),\qquad H H^{T}=\mathbb{I},
\label{eq:H-def}
\end{equation}
and therefore leaves $m_\nu$ unchanged. This observation anticipates the group-theoretic classification developed in the next section and makes explicit how the $m_\nu$-preserving freedom acts on the right of $\Omega$ (or, equivalently, of $R$) while preserving Eq.~\eqref{eq:exact-id}.

\section{Classification of the $m_\nu$-preserving freedom}
\label{sec:classification}
We classify the full set of right-multiplications on $R$ that keep $m_\nu$ invariant by introducing
\begin{equation}
\mathcal{G}\thickspace \equiv\thickspace \{\thinspace F\in GL(3,\mathbb{C})\ \big|\ F D_N F^{T}=D_N\thinspace \},
\qquad D_N=\mathrm{diag}(M_1,M_2,M_3)>0 .
\label{eq:G-def}
\end{equation}

\begin{Theorem}[Conjugate complex-orthogonal classification]
\label{thm:classification-again}
With $D_N>0$ diagonal as above,
\begin{equation}
\mathcal{G} \thickspace =\thickspace  D_N^{1/2}\thinspace O(3,\mathbb{C})\thinspace D_N^{-1/2}
\thickspace =\thickspace \{\thinspace  D_N^{1/2} H D_N^{-1/2}\thinspace :\thinspace  H\in O(3,\mathbb{C})\thinspace \},
\label{eq:G-class}
\end{equation}
where $O(3,\mathbb{C})\equiv\{H\in GL(3,\mathbb{C})\ |\ H H^{T}=\mathbb{I}\}$ is the complex orthogonal group.
\end{Theorem}

\begin{proof}
($\subseteq$) Take $F\in\mathcal{G}$ and set $H\equiv D_N^{-1/2} F D_N^{1/2}$. Then
$H H^{T}=D_N^{-1/2} (F D_N F^{T}) D_N^{-1/2}= \mathbb{I}$, so $H\in O(3,\mathbb{C})$ and $F=D_N^{1/2} H D_N^{-1/2}$.
($\supseteq$) Conversely, if $F=D_N^{1/2} H D_N^{-1/2}$ with $H H^{T}=\mathbb{I}$, then
$F D_N F^{T}=D_N^{1/2} H D_N^{-1/2} D_N D_N^{-1/2} H^{T} D_N^{1/2}= D_N^{1/2}(H H^{T})D_N^{1/2}=D_N$.
\end{proof}
We present it here in a form tailored to the seesaw problem. To the best of our knowledge, this may not constitute a new mathematical theorem. However, we could not locate a reference where this exact seesaw-tailored formulation is stated in this clear form.

Eq.~\eqref{eq:G-class} exhibits $\mathcal{G}$ as the conjugate of $O(3,\mathbb{C})$ by $D_N^{1/2}$. The map
\begin{equation}
\Phi:\ O(3,\mathbb{C}) \longrightarrow \mathcal{G},\qquad \Phi(H)=D_N^{1/2} H D_N^{-1/2},
\end{equation}
is a group isomorphism with inverse $\Phi^{-1}(F)=D_N^{-1/2} F D_N^{1/2}$. Hence $\mathcal{G}$ inherits the global structure of $O(3,\mathbb{C})$: it has two connected components labeled by $\det=\pm 1$ and complex dimension three. The Lie algebra $\mathfrak{so}(3,\mathbb{C})$ consists of complex antisymmetric matrices, so any $H\in O(3,\mathbb{C})$ can be written as $\exp(z_{12} L_{12}+z_{23} L_{23}+z_{13} L_{13})$ with $z_{ij}\in\mathbb{C}$, or as a product of plane rotations $R_{ij}(z)$ acting in $(i,j)$ blocks by $\begin{pmatrix}\cos z & \sin z\\ -\sin z & \cos z\end{pmatrix}$, using $\cos^2 z+\sin^2 z=1$ in $\mathbb{C}$. \\
We group representatives into four spectral families:
\begin{enumerate}[label=(\roman*), leftmargin=*, itemsep=2pt, topsep=4pt]
  \item \textbf{Family E (elliptic):} products of $R_{ij}(x)$ with $x\in\mathbb{R}$
  (all eigenvalues on the unit circle);
  \item \textbf{Family H (hyperbolic):} products of $R_{ij}(iy)$ with $y\in\mathbb{R}$
  (eigenvalues occur in reciprocal pairs with $|\lambda|\neq 1$);
  \item \textbf{Family P (parity-like):} $H^2=I$ (reflections, permutations such as $P_{23}$);
  \item \textbf{Family EH-mixed:} generic $R_{ij}(x+iy)$ not reducible to a single type.
\end{enumerate}

The six benchmark points used later (identity, a diagonal reflection, $P_{23}$, and three plane rotations) are simple representatives of our four spectral families: identity and real-angle rotations belong to Family $E$; reflections and $P_{23}$ to Family $P$; pure-imaginary rotations to Family $H$; and generic complex rotations to the EH-mixed family. For convenience, we label these six benchmark points by $A,\dots,F$; their explicit matrices $H$ are given in Appendix~\ref{six_classes}.

Since $\det H=\pm 1$, one has $\det F=\det H=\pm 1$. The $\det(H)=+1$ component is $SO(3,\mathbb{C})=\exp(\mathfrak{so}(3,\mathbb{C}))$, while the $\det(H)=-1$ component is obtained by multiplying any $SO(3,\mathbb{C})$ element by a reflection.

Although $m_\nu$ is invariant, the heavy–light sector is generally rotated and rescaled. From $R'\negthinspace =\negthinspace R F$ and $F\negthinspace =\negthinspace D_N^{1/2} H D_N^{-1/2}$,
\begin{equation}
\eta' \thickspace =\thickspace  \tfrac12 R' R'^\dagger \thickspace =\thickspace  \tfrac12 R (F F^\dagger) R^\dagger
\thickspace =\thickspace  \tfrac12 R \big(D_N^{1/2} H H^\dagger D_N^{1/2}\big) R^\dagger,
\label{eq:eta-transform}
\end{equation}
which equals $\eta$ only when $H$ is unitary, a measure-zero subset of $O(3,\mathbb{C})$. This guarantees class sensitivity of flavor invariants that depend on $\eta$ and $m_D^\dagger m_D$. A notable constant of motion is the determinant of $\eta$: using $R=i\thinspace N\sqrt{D_\nu}\thinspace \Omega D_N^{-1/2}$ with $\Omega\in O(3,\mathbb{C})$,
\begin{equation}
\det(\eta)=2^{-3}\det(RR^\dagger)
=2^{-3}\det(D_\nu)\det(D_N^{-1})\thinspace \det(\Omega\Omega^\dagger)
=2^{-3}\det(D_\nu)\det(D_N^{-1}),
\end{equation}
since $|\det\Omega|=1$. Thus $\det\eta$ depends only on $(D_\nu,D_N)$ and is independent of the $F$-class.

If $D_N$ has degeneracies, the stabilizer enlarges in the degenerate subspace. For instance, if $M_2=M_3$, then in the $(2,3)$ plane one may take any $H_{23}\in O(2,\mathbb{C})$, so $\mathcal{G}$ contains the conjugate of $O(2,\mathbb{C})$ acting on that plane. More generally, for a multiplicity pattern $D_N=\mathrm{diag}(M^{(1)}\mathbb{I}_{k_1}, M^{(2)}\mathbb{I}_{k_2},\ldots)$ the invariance group contains the product of conjugates of $O(k_a,\mathbb{C})$ on each degenerate block. This explains both the enhancement of freedom at degeneracy and the suppression of the unflavored CP-odd invariant $\mathcal{I}_{\rm CP}^{(1)}\propto (M_i^2-M_j^2)$ for degenerate pairs, motivating the flavored generalizations developed later. In summary, all $m_\nu$-preserving right actions are exhausted by the conjugate complex-orthogonal group in Eq.~\eqref{eq:G-class}. The freedom acts on the right of $\Omega$ or $R$, leaves $m_\nu$ invariant by construction, and generically rotates the heavy–light sector, producing the class-sensitive signals explored next.


\section{Flavor invariants: class-blind versus class-sensitive}
\label{sec:invariants}
This section develops a basis-invariant diagnostic set that separates the $m_\nu$-preserving classes introduced in Section~\ref{sec:classification}. We work with the weak-basis transformations summarized in Eq.~\eqref{eq:WBT} and use the Hermitian combinations
$h_\nu=m_\nu m_\nu^\dagger$, $H_D=m_D^\dagger m_D$, and $X=M_R^\dagger M_R$ \citep{Branco(2005),Jianlong(2022),Jianlong(2021)}. All statements below are independent of any seesaw expansion and rely only on exact tree-level matching.

\begin{Lemma}
\label{lem:mnu-similarity}
Under a weak-basis transformation with unitary $W_L,W_R$, the light-neutrino matrix transforms as
\begin{equation}
m_\nu \longrightarrow m_\nu' = W_L\thinspace m_\nu\thinspace W_L^{T},
\qquad
h_\nu \longrightarrow h_\nu' = W_L\thinspace h_\nu\thinspace W_L^\dagger .
\end{equation}
\end{Lemma}

\begin{proof}
From Eq.~\eqref{eq:WBT}, $m_D\to W_L m_D W_R^\dagger$ and $M_R^{-1}\to W_R M_R^{-1} W_R^{T}$. Therefore
$m_\nu'=-W_L m_D W_R^\dagger W_R M_R^{-1} W_R^{T} (W_L m_D W_R^\dagger)^{T}
= W_L (-m_D M_R^{-1} m_D^{T}) W_L^{T} = W_L m_\nu W_L^{T}$, and the statement for $h_\nu$ follows.
\end{proof}

\begin{Proposition}[Class-blind low-energy controls]
\label{prop:classblind-controls}
For any $k\in\mathbb{N}$, the scalars $\Tr(h_\nu^{\thinspace k})$ and $\det h_\nu$ are invariant under weak-basis transformations. Moreover, if $R' = R F$ with $F\in\mathcal{G}$, then $m_\nu'=-R' D_N R'^{T}=m_\nu$ and thus all $\Tr(h_\nu^{\thinspace k})$ and $\det h_\nu$ are identical across $F$-classes.
\end{Proposition}

\begin{proof}
Invariance under weak-basis transformations follows from Lemma~\ref{lem:mnu-similarity} by unitary similarity. For the second claim, $F\in\mathcal{G}$ implies $F D_N F^{T}=D_N$, hence $-R' D_N R'^{T}=-R D_N R^{T}=m_\nu$.
\end{proof}

To build class-sensitive invariants we need the transformation of the heavy--light block in the exact Takagi diagonalization.

\begin{Lemma}
\label{lem:R-eta-transform}
There exist unitary $V_\nu,V_N$ such that under weak-basis transformations one can choose the diagonalizer so that
\begin{equation}
R \longrightarrow R' = W_L\thinspace R\thinspace V_N^\dagger,\qquad
\eta \equiv \tfrac12 R R^\dagger \longrightarrow \eta' = W_L\thinspace \eta\thinspace W_L^\dagger .
\end{equation}
\end{Lemma}

\begin{proof}
Let $\mathcal{U}$ diagonalize $\mathcal{M}$ and define $\mathcal{U}'=\mathrm{diag}(W_L,W_R^{*})\thinspace \mathcal{U}\thinspace \mathrm{diag}(V_\nu,V_N)$. Then $\mathcal{U}'$ diagonalizes $\mathcal{M}'$ into the same eigenvalues by construction, and reading the upper-right block gives $R' = W_L R V_N^\dagger$. The statement for $\eta$ follows.
\end{proof}

\begin{Proposition}[Non-unitarity invariants]
\label{prop:nonunitarity-invariants}
The quantities
\begin{equation}
J_{\eta,1}=\Tr(\eta),\qquad
J_{\eta,2}=\Tr(\eta^2),\qquad
J_{\eta,3}=\det(\eta)
\end{equation}
are invariant under weak-basis transformations. Under the $m_\nu$-preserving right action $R\to R F$ with $F\in\mathcal{G}$ they transform to
$\eta'=\tfrac12 R (F F^\dagger) R^\dagger$ and are therefore class-sensitive in general (i.e.\ they change with $F$), except for $J_{\eta,3}$ which is constant across the entire class, as shown in Proposition~\ref{prop:det-eta-constant}.
\end{Proposition}

\begin{proof}
Weak-basis invariance follows from Lemma~\ref{lem:R-eta-transform} by similarity. The $F$-dependence is immediate from the displayed transformation. That $J_{\eta,3}$ is actually $F$-independent is proved below.
\end{proof}

\begin{Proposition}[Alignment invariant]
\label{prop:alignment}
The scalar
\begin{equation}
K_{\rm align} \thickspace \equiv\thickspace  \Tr\big([\eta,h_\nu]^2\big),
\qquad [A,B]=AB-BA,
\end{equation}
is invariant under weak-basis transformations. If $\eta$ and $h_\nu$ are Hermitian, then $K_{\rm align}\in\mathbb{R}_{\le 0}$ and $K_{\rm align}=0$ if and only if $[\eta,h_\nu]=0$.
\end{Proposition}

\begin{proof}
Under $W_L$, both $\eta$ and $h_\nu$ transform by similarity, hence so does their commutator, and the trace of its square is invariant. For Hermitian $\eta,h_\nu$, the commutator is anti-Hermitian, $C^\dagger=-C$. Writing $C=iK$ with Hermitian $K$, $\Tr(C^2)=-\Tr(K^2)\le 0$, and it vanishes iff $K=0$ iff $C=0$.
\end{proof}

\begin{Theorem}[A CP-odd leptogenesis invariant]
\label{thm:cpodd-invariant}
Let $H_D=m_D^\dagger m_D$ and $X=M_R^\dagger M_R$. The quantity
\begin{equation}
\mathcal{I}_{\rm CP}^{(1)} \thickspace \equiv\thickspace  \Im\thinspace \Tr\negthinspace \big( H_D\thinspace X^{1/2}\thinspace H_D^{T}\thinspace X^{3/2} \big)
\label{eq:ICP-def}
\end{equation}
is invariant under weak-basis transformations. In the basis where $M_R=D_N=\mathrm{diag}(M_1,M_2,M_3)>0$ it reduces to
\begin{equation}
\mathcal{I}_{\rm CP}^{(1)} \thickspace =\thickspace  \sum_{i<j} (M_i^2-M_j^2)\thinspace M_i M_j\thickspace \Im\negthinspace \Big[(H_D)_{ij}^{\thinspace 2}\Big] .
\label{eq:ICP-mass-basis}
\end{equation}
\end{Theorem}

\begin{proof}
Under weak-basis transformations $H_D\to W_R H_D W_R^\dagger$, $X\to W_R X W_R^\dagger$, and $H_D^{T}\to W_R^{*} H_D^{T} W_R^{T}$. Functional calculus implies $X^p\to W_R X^p W_R^\dagger$ for any real $p$. Cyclicity of the trace cancels all $W_R$ factors, proving invariance. In the $M_R$-diagonal basis, writing out the trace gives
$\Im\sum_{i,j} X_i^{1/2} X_j^{3/2} (H_D)_{ij}(H_D)_{ji}=\sum_{i<j}(X_j - X_i)\sqrt{X_i X_j}\thinspace \Im[(H_D)_{ij}^{2}]$ with $X_i=M_i^2$, which is Eq.~\eqref{eq:ICP-mass-basis}.
\end{proof}

\begin{Proposition}[Class-independence of $\det\eta$]
\label{prop:det-eta-constant}
For fixed $(U,D_\nu,D_N)$ and any $F\in\mathcal{G}$, the determinant of the non-unitarity matrix is
\begin{equation}
\det\eta \thickspace =\thickspace  2^{-3}\thinspace \det(D_\nu)\thinspace \det(D_N^{-1}),
\label{eq:deteta-constant}
\end{equation}
independent of the $F$-class.
\end{Proposition}

\begin{proof}
Using the Casas--Ibarra parameterization $m_D=i\thinspace U\sqrt{D_\nu}\thinspace \Omega\thinspace \sqrt{D_N}$ with $\Omega\in O(3,\mathbb{C})$, the exact heavy--light block is $R=i\thinspace U\sqrt{D_\nu}\thinspace \Omega\thinspace D_N^{-1/2}$. Hence
\begin{align}
\det(\eta)
=2^{-3}\det(RR^\dagger)
&=2^{-3}\det\negthinspace \big(U\sqrt{D_\nu}\thinspace \Omega\thinspace D_N^{-1/2} D_N^{-1/2}\thinspace \Omega^\dagger \sqrt{D_\nu}\thinspace U^\dagger\big) \nonumber\\
&=2^{-3}\det(D_\nu)\thinspace \det(D_N^{-1})\thinspace \det(\Omega\Omega^\dagger)\nonumber\\
&=2^{-3}\det(D_\nu)\thinspace \det(D_N^{-1}),
\end{align}
since $\det U$ is unimodular and $|\det\Omega|=1$ for $\Omega\in O(3,\mathbb{C})$. Right-multiplying $R$ by any $F\in\mathcal{G}$ corresponds to $\Omega\to\Omega H$ with $H\in O(3,\mathbb{C})$, which preserves $|\det\Omega|$ and thus leaves $\det\eta$ unchanged.
\end{proof}

The above results divide flavor invariants into two families. Under the right action $R\to R F$ with $F\in\mathcal{G}$, the flavor invariants split into two groups:
\begin{itemize}[leftmargin=*,itemsep=2pt]
  \item \textbf{Class-blind (unchanged under $F$):}
  \begin{itemize}[leftmargin=1.4em,itemsep=1pt]
    \item spectrum of $h_\nu\equiv m_\nu m_\nu^\dagger$ (equivalently $D_\nu$);
    \item $\mathrm{Tr}\,h_\nu$, $\mathrm{Tr}\,h_\nu^{2}$, $\det h_\nu$;
    \item $\det\eta$ with $\eta\equiv \tfrac12 RR^\dagger$; in our setup $\det\eta = 2^{-3}\,\det D_\nu\,\det D_N^{-1}$, fixed entirely by $(D_\nu,D_N)$.
  \end{itemize}
  \item \textbf{Class-sensitive (change with $F$):}
  \begin{itemize}[leftmargin=1.4em,itemsep=1pt]
    \item $\mathrm{Tr}\,\eta$, $\mathrm{Tr}\,\eta^{2}$ (equivalently the eigenvalues of $\eta$);
    \item $K_{\rm align}\equiv \mathrm{Tr}\!\left([\eta,h_\nu]^2\right)$;
    \item CP-odd traces such as $\mathcal{I}^{(1)}_{\rm CP}$ and flavored extensions.
  \end{itemize}
\end{itemize}


\section{Six benchmark points and numerical fingerprints}
\label{sec:benchmark}
To illustrate the classification and the invariant diagnostics, we keep $(U,D_\nu,D_N)$ fixed and scan six representative elements of the $m_\nu$-preserving group $\mathcal{G}=D_N^{1/2}O(3,\mathbb{C})D_N^{-1/2}$. Throughout this section we take
\begin{equation}
D_\nu = \mathrm{diag}(0.001, 0.00866, 0.050)~\mathrm{eV},\qquad
D_N^{\mathrm{(high)}} = \mathrm{diag}(3,5,8)\times 10^{11}~\mathrm{GeV},
\label{eq:benchmark-Dn}
\end{equation}
and a PMNS matrix with $\theta_{23}=45^\circ$ and $\delta = -\pi/2$ (Majorana phases set
to zero for definiteness), so that $|U_{\mu i}| = |U_{\tau i}|$ holds exactly. These choices
are representative of current practice: $\theta_{23} = 45^\circ$ as an octant-symmetric
midpoint~\cite{Esteban(2024)}, $m_1 = 10^{-3}\,\mathrm{eV}$ as a convenient normal-ordering scale~\cite{PDG(2024)},
and $D_N^{\mathrm{(high)}}$ as a typical high-scale type-I seesaw spectrum consistent with
thermal leptogenesis~\cite{Davidson(2008)}. \\
In addition, to illustrate that the invariant fingerprints are scale-agnostic and to provide
an explicit GeV-scale realization, we consider a rescaled low-scale benchmark
\begin{equation}
D_N^{\mathrm{(low)}} = \mathrm{diag}(3,5,8)~\mathrm{GeV}
= 10^{-11}\, D_N^{\mathrm{(high)}},
\end{equation}
keeping $(U_{\mathrm{PMNS}}, D_\nu)$ and the Casas–Ibarra matrix $\Omega$ fixed. These
two benchmarks are sufficient to exhibit all structural features we emphasize: class-blind
quantities remain unchanged, while the class-sensitive diagnostics
$(\mathrm{Tr}\,\eta,\mathrm{Tr}\,\eta^2, K_{\mathrm{align}}, \mathcal{I}_{\mathrm{CP}}^{(1)})$ split across the
$F$-classes and scale predictably with $D_N$. We now select six specific choices of $\Omega$ (equivalently $H$), which we
treat as numerical benchmark points and label by $A$--$F$; the corresponding
$H$ matrices are listed explicitly in Appendix~\ref{six_classes}, and each point lies in a
distinct $F$-class in the classification of Section~\ref{sec:classification}.

The Dirac mass is parameterized in Casas–Ibarra form \cite{Casas(2001)},
\begin{equation}
m_D = i\thinspace N\thinspace \sqrt{D_\nu}\thinspace \Omega\thinspace \sqrt{D_N},\qquad \Omega\in O(3,\mathbb{C}),
\end{equation}
and the exact heavy–light block is
\begin{equation}
R = i\thinspace N\thinspace \sqrt{D_\nu}\thinspace \Omega\thinspace D_N^{-1/2},\qquad
m_\nu=-R D_N R^{T}.
\end{equation}
Right-multiplying $R$ by $F\in\mathcal{G}$ is equivalent to acting on the right of $\Omega$ by
$H=D_N^{-1/2} F D_N^{1/2}\in O(3,\mathbb{C})$, namely $\Omega\to \Omega H$. By construction, $m_\nu$ remains unchanged.

We use six concrete choices of $H$ that exemplify reflections, permutations, and complexified plane rotations. Writing
$\rho_{ij}\equiv\sqrt{M_i/M_j}$ and denoting by $R_{ij}(z)$ the plane rotation acting as $\begin{pmatrix}\cos z & \sin z\\ -\sin z & \cos z\end{pmatrix}$ in the $(i,j)$ block (with $z\in\mathbb{C}$), we take
\begin{itemize}
\item \text{Identity (Family E; $\det H=+1$)}:\\ $H=I$,\qquad\ \ \thinspace  $F=I$;\\
    \item \text{Diagonal reflection (Family P, commuting with $D_N$; $\det H=-1$)}:\\ $H=\mathrm{diag}(-1,\thinspace 1,\thinspace 1),\quad F=\mathrm{diag}(-1,\thinspace 1,\thinspace 1)$;\\
\item \text{Permutation $P_{23}$ (Family P, non-commuting with $D_N$; $\det H=-1$)}:\\ $H=P_{23},\quad
F=\begin{pmatrix}
1&0&0\\[2pt]
0&0&\rho_{23}\\[2pt]
0&\rho_{32}&0
\end{pmatrix}$;\\
\item \text{Pure imaginary rotation $R_{12}(0.7i)$ (Family H; boost in $12$-plane)}:\\ $H=R_{12}(0.7\thinspace i),\quad
F_{(1,2)}=\begin{pmatrix}
\cosh 0.7 & i\thinspace \sinh 0.7\thinspace \rho_{12}\\[2pt]
-\thinspace i\thinspace \sinh 0.7\thinspace \rho_{21} & \cosh 0.7
\end{pmatrix}$;\\
\item \text{Complex rotation $R_{23}(0.5+0.3i)$ (Family EH; mixed)}:\\ $H=R_{23}(0.5+0.3\thinspace i),\quad
F_{(2,3)}=\begin{pmatrix}
c & s\thinspace \rho_{23}\\[2pt]
-\thinspace s\thinspace \rho_{32} & c
\end{pmatrix},\ \ c=\cos(0.5+0.3i),\ s=\sin(0.5+0.3i)$;\\
\item \text{Real rotation $R_{13}(0.9)$ (Family E; elliptic in $13$-plane; $\det H=+1$)}:\\ $H=R_{13}(0.9),\quad
F_{(1,3)}=\begin{pmatrix}
\cos 0.9 & \sin 0.9\thinspace \rho_{13}\\[2pt]
-\thinspace \sin 0.9\thinspace \rho_{31} & \cos 0.9
\end{pmatrix}$.
\end{itemize} 
These six elements are not intended to be exhaustive. They simply provide a convenient set of benchmarks spanning the two components with $\det H=\pm1$ and sampling both real and imaginary “angles.” For each benchmark point $A$--$F$ we evaluate the flavor-invariant controls built solely from $h_\nu=m_\nu m_\nu^\dagger$,
\begin{equation}
I_{\nu,1}=\Tr(h_\nu),\qquad I_{\nu,2}=\Tr(h_\nu^2),\qquad I_{\nu,3}=\det h_\nu,
\end{equation}
and the class-sensitive set
\begin{equation}
J_{\eta,1}=\Tr(\eta),\qquad J_{\eta,2}=\Tr(\eta^2),\qquad
K_{\rm align}=\Tr\big([\eta,h_\nu]^2\big),\qquad
\mathcal{I}_{\rm CP}^{(1)}=\Im\thinspace \Tr\negthinspace \big(H_D X^{1/2} H_D^{T} X^{3/2}\big),
\end{equation}
with $\eta=\tfrac12 R R^\dagger$, $H_D=m_D^\dagger m_D$, and $X=M_R^\dagger M_R$. 

By Proposition~\ref{prop:det-eta-constant}, $\det\eta=2^{-3}\det(D_\nu)\det(D_N^{-1})$ is the same for all $F$-classes, so all six benchmark points $A$--$F$ share the same value
and it is therefore not repeated in the tables. The class-blind controls $(I_{\nu,1},I_{\nu,2},I_{\nu,3})$
are likewise identical for all six benchmark points.\\
For compactness we tabulate the class-sensitive fingerprints for the six
benchmark points $A$--$F$ in Table~\ref{tab:highscale} (high-scale) and Table~\ref{tab:lowscale} (low-scale). At the high-scale
benchmark $D_N^{\mathrm{(high)}}$ in Eq.~(\ref{eq:benchmark-Dn}), the values of
$\mathrm{Tr}\,\eta$, $\mathrm{Tr}\,\eta^2$, $K_{\mathrm{align}}$ and $I_{\mathrm{CP}}^{(1)}$ for the six
benchmark points $A$–$F$ are collected in Table~\ref{tab:highscale}. To make contact with GeV-scale
realizations, we then repeat the computation at the low-scale benchmark
$D_N^{\mathrm{(low)}} = \mathrm{diag}(3,5,8)\,\mathrm{GeV}$, keeping $(U_{\mathrm{PMNS}}, D_\nu, H)$
fixed. The corresponding fingerprints are shown in Table~\ref{tab:lowscale}. As anticipated from
$R \propto D_N^{-1/2}$, one observes the scaling
$\mathrm{Tr}\,\eta \propto D_N^{-1}$ and
$\mathrm{Tr}\,\eta^2,\,K_{\mathrm{align}} \propto D_N^{-2}$, while the relative pattern across
benchmarks $A$–$F$ is unchanged.\\
The numerical patterns mirror the analytic expectations of Section~\ref{sec:invariants}: $J_{\eta,1}$ and
$J_{\eta,2}$ cluster within factors of order unity, $K_{\mathrm{align}}$ spreads over many decades,
and $I_{\mathrm{CP}}^{(1)}$ may flip sign across benchmarks.

\begin{table*}[t]
\centering
\caption{Class–sensitive invariant fingerprints for six benchmark points $A$–$F$ representing distinct $F$–classes at fixed low–energy data
$(U_{\rm PMNS}, D_\nu)$ and a \textbf{high–scale} heavy–neutrino spectrum
$D_N=D_N^{(\text{high})} = \mathrm{diag}(3,5,8)\times 10^{11}\,\mathrm{GeV}$.
Entries are computed with
$\theta_{12}=33.44^\circ$, $\theta_{13}=8.57^\circ$, $\theta_{23}=45^\circ$,
$\delta=-\pi/2$, vanishing Majorana phases, and $\Omega = H$.
Units: $\Tr\eta$ and $\Tr\eta^2$ are dimensionless;
$K_{\rm align}$ is in $\mathrm{GeV}^4$;
$\mathcal{I}^{(1)}_{\rm CP}$ is in $\mathrm{GeV}^8$.
The constant
$\det\eta = 2^{-3}\det(D_\nu)\det(D_N^{-1})
\simeq 4.5104\times 10^{-70}$ (dimensionless)
is identical for all benchmark points.
Numerical details are given in Appendix~\ref{app:details}.}
\label{tab:highscale}
\begin{tabular}{lcccc}
\toprule
Benchmark & $\Tr(\eta)$ & $\Tr(\eta^2)$ & $K_{\rm align}=\Tr([\eta,h_\nu]^2)$ & $\mathcal{I}_{\rm CP}^{(1)}$\\
\midrule
A & $4.158\times10^{-23}$ & $1.054\times10^{-45}$ & $-3.789\times10^{-109}$ & $+1.615\times10^{39}$\\
B & $4.158\times10^{-23}$ & $1.054\times10^{-45}$ & $-3.789\times10^{-109}$ & $-1.210\times10^{39}$\\
C & $5.708\times10^{-23}$ & $2.532\times10^{-45}$ & $-5.101\times10^{-109}$ & $-1.822\times10^{37}$\\
D & $5.640\times10^{-23}$ & $1.580\times10^{-45}$ & $-6.114\times10^{-91}$  & $-7.760\times10^{47}$\\
E & $5.398\times10^{-23}$ & $2.198\times10^{-45}$ & $-1.489\times10^{-87}$  & $-8.945\times10^{49}$\\
F & $7.290\times10^{-23}$ & $4.097\times10^{-45}$ & $-1.607\times10^{-88}$  & $+9.785\times10^{38}$\\
\bottomrule
\end{tabular}
\end{table*}

All numerical computations were performed in IEEE 754 double-precision floating-point arithmetic (float64 / complex128 in NumPy), without
relying on a series expansion in $m_D/M_R$. We verified numerically that $m_\nu=-R D_N R^{T}$ holds to machine precision for each benchmark
point $A$--$F$ and that the weak-basis invariant statements of Section~\ref{sec:invariants} are respected. All numerical results reported in Tables~\ref{tab:highscale}
and~\ref{tab:lowscale} were generated with the author's publicly
available Python code~\cite{LuCodeRepo}.

\begin{table*}[t]
\centering
\caption{Same as Table~\ref{tab:highscale} but for a low–scale heavy–neutrino
spectrum $D_N = D_N^{(\text{low})}= \mathrm{diag}(3,5,8)\,\mathrm{GeV}$,
obtained from the high–scale benchmark by an overall rescaling
$D_N \to 10^{-11} D_N$ at fixed $(U_{\rm PMNS}, D_\nu)$ and $H$.
As expected from the scaling $R\propto D_N^{-1/2}$,
the class–sensitive invariants obey
$\Tr\eta \propto D_N^{-1}$,
$\Tr\eta^2,\,K_{\rm align}\propto D_N^{-2}$,
while the relative pattern across $A$–$F$ is unchanged.
Units are as in Table~\ref{tab:highscale}.
The constant
$\det\eta = 2^{-3}\det(D_\nu)\det(D_N^{-1})
\simeq 4.5104\times 10^{-37}$ (dimensionless)
is again identical for all benchmark points.}
\label{tab:lowscale}
\begin{tabular}{lcccc}
\toprule
Benchmark & $\Tr(\eta)$ & $\Tr(\eta^2)$ & $K_{\rm align}=\Tr([\eta,h_\nu]^2)$ & $\mathcal{I}_{\rm CP}^{(1)}$\\
\midrule
A & $4.158\times10^{-12}$ & $1.054\times10^{-23}$ & $-3.789\times10^{-87}$ & $+1.615\times10^{-27}$\\
B & $4.158\times10^{-12}$ & $1.054\times10^{-23}$ & $-3.789\times10^{-87}$ & $-1.210\times10^{-27}$\\
C & $5.708\times10^{-12}$ & $2.532\times10^{-23}$ & $-5.102\times10^{-87}$ & $-1.822\times10^{-29}$\\
D & $5.640\times10^{-12}$ & $1.580\times10^{-23}$ & $-6.114\times10^{-69}$  & $-7.760\times10^{-19}$\\
E & $5.398\times10^{-12}$ & $2.198\times10^{-23}$ & $-1.489\times10^{-65}$  & $-8.945\times10^{-17}$\\
F & $7.290\times10^{-12}$ & $4.097\times10^{-23}$ & $-1.607\times10^{-66}$  & $+9.785\times10^{-28}$\\
\bottomrule
\end{tabular}
\end{table*}


\section{Heavy-mass degeneracies and flavor invariants}
\label{sec:degeneracy}
When two or more heavy eigenvalues coincide, the $m_\nu$-preserving freedom enlarges inside the degenerate subspace. If, for instance, $M_2=M_3$, then any transformation acting as $H_{23}\in O(2,\mathbb{C})$ on the $(2,3)$ plane leaves $D_N$ invariant and therefore belongs to $\mathcal{G}=D_N^{1/2}O(3,\mathbb{C})D_N^{-1/2}$. More generally, for a multiplicity pattern
\begin{equation}
D_N=\mathrm{diag}\negthinspace \big(M^{(1)}\mathbb{I}_{k_1},\thinspace M^{(2)}\mathbb{I}_{k_2},\thinspace \ldots\big),
\end{equation}
the invariance group contains the product of conjugates of $O(k_a,\mathbb{C})$ on each block, with the non-degenerate directions transforming by isolated reflections/rotations. This explains the enhancement of class freedom at degeneracy mentioned earlier.

The unflavored CP-odd invariant of Theorem~\ref{thm:cpodd-invariant} immediately shows how degeneracy suppresses CP probes that are antisymmetric in masses. In the $M_R$-diagonal basis,
\begin{equation}
\mathcal{I}_{\rm CP}^{(1)}=\sum_{i<j}(M_i^2-M_j^2)\thinspace M_i M_j\thickspace \Im\negthinspace \big[(H_D)_{ij}^{\thinspace 2}\big],
\end{equation}
so any exactly degenerate pair contributes zero by construction. Near-degenerate pairs are accordingly suppressed by the small mass splitting, thereby hiding part of the class sensitivity in the unflavored sector. This motivates the use of flavor invariants which remain sensitive to the orientation of the Dirac structures even when two heavy masses coincide.

To construct flavored, weak-basis invariant probes, it is convenient to introduce the rank-one projectors onto charged-lepton flavors in the charged-lepton mass basis,
\begin{equation}
\Pi_\alpha = \mathrm{diag}(\delta_{e\alpha},\delta_{\mu\alpha},\delta_{\tau\alpha}),\qquad \alpha=e,\mu,\tau.
\end{equation}
Under weak-basis transformations $W_L\in U(3)$ acting on $\ell_L$, these projectors transform covariantly, $\Pi_\alpha\to W_L \Pi_\alpha W_L^\dagger$, so traces built by inserting $\Pi_\alpha$ remain basis invariant by cyclicity. A simple family of flavored CP-odd invariants is then
\begin{equation}
\mathcal{I}^{(p,q)}_{{\rm CP},\alpha}\thickspace \equiv\thickspace 
\Im\thinspace \Tr\negthinspace \Big(\Pi_\alpha\thinspace m_D\thinspace X^{p}\thinspace m_D^\dagger\thinspace \Pi_\alpha\thinspace m_D\thinspace X^{q}\thinspace m_D^\dagger\Big),
\qquad p,q\in\mathbb{R},
\label{eq:flavored-family}
\end{equation}
with $X\equiv M_R^\dagger M_R$. Invariance under weak-basis transformations follows from $m_D\to W_L m_D W_R^\dagger$, $X\to W_R X W_R^\dagger$, $\Pi_\alpha\to W_L \Pi_\alpha W_L^\dagger$, and cyclicity of the trace which removes both $W_L$ and $W_R$. In the $M_R$-diagonal basis one obtains the mass-explicit form
\begin{equation}
\mathcal{I}^{(p,q)}_{{\rm CP},\alpha}
=\sum_{i<j}\big(M_i^{2p} M_j^{2q}-M_i^{2q} M_j^{2p}\big)\thickspace 
\Im\negthinspace \Big[(m_D)_{\alpha i}\thinspace (m_D^\dagger m_D)_{ij}\thinspace (m_D^\dagger)_{j\alpha}\Big].
\label{eq:flavored-mass-basis}
\end{equation}
While the coefficient in parentheses again vanishes at exact degeneracy $M_i=M_j$, different choices of $(p,q)$ control the hierarchy of suppression near degeneracy and allow one to optimize sensitivity to specific heavy scales. In particular, $(p,q)=(0,1)$ or $(\tfrac12,\tfrac32)$ reproduce the overall mass scalings familiar from unflavored leptogenesis at leading order, now with flavor tags carried by $\Pi_\alpha$.

Strict equality of heavy masses represents a limit in which resonant effects become important and the standard “mass-difference” invariants are driven to zero at fixed order. In that regime, physical CP asymmetries are regulated by width effects that can be organized in a basis-invariant way using commutators of $X$ with the width matrix $\Gamma\propto H_D/M_R$ (schematically, $[X,\Gamma]\neq 0$). While a full treatment of resonant leptogenesis is beyond our scope, we note that one can define regulated, flavored CP-odd traces by the analytic continuation $X\to X\pm i\thinspace \Gamma$ inside Eq.~\eqref{eq:flavored-family} so that the antisymmetric mass differences are replaced by combinations of complex eigenvalues, thereby avoiding the artificial zero at exact degeneracy. Independently of the regulator, the non-unitarity sector remains a powerful discriminator: the invariants
\begin{equation}
\Tr(\eta),\qquad \Tr(\eta^2),\qquad K_{\rm align}=\Tr\big([\eta,h_\nu]^2\big)
\end{equation}
are untouched by mass degeneracy in $D_N$ and retain their strong class sensitivity through the orientation and rescaling encoded in $F F^\dagger$.

To summarize, heavy-mass degeneracies enlarge the $m_\nu$-preserving symmetry inside the degenerate subspace and suppress unflavored CP-odd probes that are explicitly antisymmetric in $(M_i^2-M_j^2)$. Flavored, weak-basis invariant constructions such as Eq.~\eqref{eq:flavored-family} restore sensitivity to the Dirac-sector orientation and, together with non-unitarity invariants, provide robust handles to separate classes even when parts of the spectrum are (nearly) degenerate. In applications to resonant leptogenesis, width effects can be incorporated via regulated, basis-invariant traces built from $X$ and $\Gamma$, ensuring that class sensitivity is maintained in the degenerate limit.


\section{Phenomenological implications}
\label{sec:pheno}
The classification $\mathcal{G}=D_N^{1/2}O(3,\mathbb{C})D_N^{-1/2}$ shows that oscillation data alone cannot distinguish between physically inequivalent completions that share the same $m_\nu$. The diagnostics built in Section~\ref{sec:invariants} identify the relevant handles and suggest a concrete program to confront classes with data. Because all scalars built from $h_\nu$ are class-blind, neutrinoless double beta decay in its standard light-neutrino exchange limit, which depends on $m_{\beta\beta}=|(m_\nu)_{ee}|$, cannot separate classes either \citep{Dolinski(2019)}. Class sensitivity instead enters through the heavy–light and Dirac sectors, i.e.\ through $\eta=\tfrac12 R R^\dagger$ and $H_D=m_D^\dagger m_D$, which control non-unitarity \cite{Aloni(2024)}, charged-lepton flavor violation \cite{Davidson(2022)}, and leptogenesis \cite{Davidson(2008)}.

Precision tests of non-unitarity constrain the Hermitian matrix $\eta$ through deviations in weak processes and neutrino production/detection \cite{Aloni(2024)}. While experimental fits are usually reported as limits on individual entries or eigenvalues of $\eta$, our invariant basis provides compact global measures. In particular,
\begin{equation}
J_{\eta,1}=\Tr\eta=\sum_{i}\lambda_i,\qquad
J_{\eta,2}=\Tr\eta^2=\sum_{i}\lambda_i^2,
\end{equation}
bound the sum and the quadratic sum of the eigenvalues $\{\lambda_i\}$ of $\eta$. The alignment invariant
\begin{equation}
K_{\rm align}=\Tr\big([\eta,h_\nu]^2\big)
\end{equation}
packs orientation information. In the eigenbasis of $h_\nu$ with $h_\nu=\mathrm{diag}(h_1,h_2,h_3)$, one finds the exact identity
\begin{equation}
K_{\rm align} \thickspace =\thickspace  -\thinspace 2 \sum_{i<j} (h_i-h_j)^2\thinspace \big|(\eta')_{ij}\big|^2,
\qquad \eta' \equiv U^\dagger \eta\thinspace  U,
\label{eq:Kalign-eig}
\end{equation}
where $U$ diagonalizes $m_\nu$ by $U^T m_\nu U=D_\nu$. Thus $K_{\rm align}$ is directly driven by the off-diagonal components of $\eta$ in the $h_\nu$ basis, weighted by the light-sector splittings. Current or future bounds on the pattern of $\eta$ can therefore be translated into constraints on $K_{\rm align}$ via Eq.~\eqref{eq:Kalign-eig}, offering a basis-invariant way to compare different analyses and experiments.

Radiative LFV decays such as $\ell_\alpha\to\ell_\beta\gamma$ and related processes constrain off-diagonal combinations of heavy–light mixing \cite{Davidson(2022)}. In the minimal Type-I seesaw and for $M_i^2\gg M_W^2$, the loop functions approach constants, so the amplitudes scale as
\begin{equation}
\mathcal{A}(\ell_\alpha\to\ell_\beta\gamma)\ \propto\ \sum_i R_{\alpha i} R_{\beta i}^* \thickspace =\thickspace  (R R^\dagger)_{\alpha\beta}\thickspace =\thickspace 2\thinspace \eta_{\alpha\beta},
\end{equation}
and the branching ratios behave approximately as $\mathrm{BR}(\ell_\alpha\to\ell_\beta\gamma)\propto |\eta_{\alpha\beta}|^2$ up to known kinematic and gauge factors. Consequently, experimental upper limits translate directly into bounds on particular directions in $\eta$-space (off-diagonal entries), complementary to the global measures $J_{\eta,1}$ and $J_{\eta,2}$. Combining both types of information restricts the allowed region of the fingerprint $(\Tr\eta,\Tr\eta^2,K_{\rm align})$ at fixed $(U,D_\nu,D_N)$ and thereby carves out admissible subsets of $H\in O(3,\mathbb{C})$.

Baryogenesis via leptogenesis depends on CP-odd rephasing invariants constructed from $H_D$ and $D_N$. In the hierarchical regime and in the $M_R$-diagonal basis, the unflavored asymmetries are proportional to $\Im[(H_D)_{ij}^2]$ weighted by mass-dependent coefficients. The invariant
\begin{equation}
\mathcal{I}_{\rm CP}^{(1)} \thickspace =\thickspace  \sum_{i<j} (M_i^2-M_j^2)\thinspace M_i M_j\thickspace \Im\negthinspace \big[(H_D)_{ij}^{\thinspace 2}\big]
\end{equation}
captures this structure and is weak-basis invariant. Its sign tracks the net sign of the summed asymmetries in the strongly hierarchical limit, while its magnitude correlates with the overall size of CP violation modulo efficiency factors. Because the right action by $H\in O(3,\mathbb{C})$ changes $H_D$ while preserving $m_\nu$, $\mathcal{I}_{\rm CP}^{(1)}$ is generically class-sensitive, as observed in the benchmark where its sign can even flip across classes. Near heavy-mass degeneracy, the unflavor invariant is suppressed by construction. In that regime, the flavor invariants introduced in Section~\ref{sec:degeneracy} maintain sensitivity and should be used instead.

These considerations suggest a practical workflow to confront the classification with data at fixed $(U,D_\nu,D_N)$: (i) sample $H\in O(3,\mathbb{C})$ (e.g.\ by products of $R_{ij}(z_{ij})$ with complex angles) and construct $F=D_N^{1/2} H D_N^{-1/2}$; (ii) compute the fingerprint
\begin{equation}
\mathbf{f}(H)\thickspace =\thickspace \big(\Tr\eta(H),\ \Tr\eta(H)^2,\ K_{\rm align}(H),\ \mathcal{I}_{\rm CP}^{(1)}(H)\big),
\end{equation}
together with individual entries $\eta_{\alpha\beta}(H)$ relevant for LFV; (iii) impose experimental bounds on non-unitarity and LFV to carve out the allowed region in $H$-space; (iv) within the surviving region, assess the range and sign of $\mathcal{I}_{\rm CP}^{(1)}$ (or its flavored analogs) consistent with successful leptogenesis for the chosen $D_N$. This procedure turns the qualitative degeneracy at fixed $m_\nu$ into a quantitative, basis-invariant map from data to theory space.

Finally, we note that while the standard light-neutrino contribution to neutrinoless double beta decay is class-blind, potential heavy-neutrino exchange amplitudes scale as $A_{0\nu\beta\beta}^{(N)}\propto \sum_i R_{e i}^2/M_i$ in the minimal seesaw and are therefore class-sensitive through $R$. In the simplest decoupling regime these contributions are typically suppressed, but in scenarios with comparatively low $M_i$ or extended dynamics they can provide an additional, complementary probe of the heavy–light sector and thus of the $F$-class.


\section{Conclusions and outlook}
\label{sec:conclusions}
We have shown that the entire $m_\nu$-preserving freedom in the canonical seesaw is exhausted by the conjugate complex-orthogonal group $\mathcal{G}=D_N^{1/2}O(3,\mathbb{C})D_N^{-1/2}$, acting on the right of the exact heavy–light block $R$ (or, equivalently, on the right of the Casas–Ibarra matrix $\Omega$) while leaving $m_\nu=-R D_N R^{T}$ unchanged. This classification promotes previously constructed examples to representatives inside a continuous symmetry manifold and makes transparent how the freedom enlarges in degenerate heavy-mass limits. Our classification relies only on the algebraic properties of the tree-level mass matrix $\mathcal{M}$ and on its exact Takagi diagonalization; no further series expansion in $m_D/M_R$ is required. Loop corrections and higher-dimensional operators are neglected throughout.

On the observable side, we separated weak-basis invariants into a class-blind family—those built solely from $h_\nu=m_\nu m_\nu^\dagger$, such as $\Tr h_\nu$, $\Tr h_\nu^2$, and $\det h_\nu$—and a class-sensitive family probing the heavy–light and Dirac sectors. The non-unitarity invariants $\Tr\eta$ and $\Tr\eta^2$, the alignment measure $K_{\rm align}=\Tr([\eta,h_\nu]^2)$, and the CP-odd leptogenesis invariant $\mathcal{I}_{\rm CP}^{(1)}$ are all weak-basis invariant yet depend on the $F$-class, with the single exception that $\det\eta$ is fixed by $(D_\nu,D_N)$, cf.\ Eq.~\eqref{eq:deteta-constant}. Analytically, $\eta\to \tfrac12 R(F F^\dagger)R^\dagger$ under $R\to R F$ ensures class sensitivity whenever $H=D_N^{-1/2}FD_N^{1/2}$ is non-unitary. Numerically, our six-class benchmark illustrates that the class-sensitive invariants can separate completions by factors of a few up to many orders of magnitude and that $\mathcal{I}_{\rm CP}^{(1)}$ may even flip sign across classes, all while the class-blind controls remain identical.

Heavy-mass degeneracies enlarge the stabilizer inside the degenerate subspace and suppress unflavored CP-odd traces that are antisymmetric in $(M_i^2-M_j^2)$. This motivates flavored, weak-basis invariant generalizations that retain sensitivity in the degenerate limit. Independently of degeneracy, the non-unitarity sector provides robust discrimination, since $F F^\dagger$ reshapes the magnitude and orientation of $\eta$ relative to $h_\nu$ and is directly testable in precision electroweak and neutrino processes.

The framework turns the qualitative degeneracy at fixed $m_\nu$ into quantitative, basis-invariant fingerprints in the space spanned by $(\Tr\eta,\Tr\eta^2,K_{\rm align},\mathcal{I}_{\rm CP}^{(1)})$ at fixed $(U_{\rm PMNS},D_\nu,D_N)$. This immediately suggests a practical program: sample $H\in O(3,\mathbb{C})$, compute the fingerprint and relevant entries of $\eta$, impose bounds from non-unitarity and LFV, and assess the size and sign of $\mathcal{I}_{\rm CP}^{(1)}$ (or flavored analogs) compatible with successful leptogenesis. The same map supports future data-driven constraints, including prospective improvements in non-unitarity searches, LFV limits, and dedicated leptogenesis studies.

Several extensions are natural. A systematic construction of flavored CP invariants near degeneracy, including width effects appropriate for resonant leptogenesis, would refine the present probes. Global analyses can translate experimental limits into exclusion regions in $H$-space, highlighting which sectors of $O(3,\mathbb{C})$ remain viable for given $(U_{\rm PMNS},D_\nu,D_N)$. Finally, renormalization-group stability of the data-favored $\mu$--$\tau$ modulus relation and its interplay with the $D_N$-orthogonal freedom may reveal infrared selections among classes. Taken together, these directions would further consolidate the invariant picture of how distinct seesaw completions populate the same low-energy neutrino physics while differing in testable heavy–light and CP properties.

Although the analytic identities derived here are fully scale–agnostic, our explicit
numerical benchmarks are specified in terms of a particular high–scale spectrum
$D_N^{\mathrm{(high)}} = \mathrm{diag}(3,5,8)\times 10^{11}\,\mathrm{GeV}$ and its
rescaled low–scale counterpart $D_N^{\mathrm{(low)}} = \mathrm{diag}(3,5,8)\,\mathrm{GeV}
=10^{-11}D_N^{\mathrm{(high)}}$. At fixed $(U_{\mathrm{PMNS}}, D_\nu)$ and Casas–Ibarra
matrix $\Omega$, one has
\begin{align}
\eta = \frac12\, U_{\mathrm{PMNS}}\sqrt{D_\nu}\,\Omega\,D_N^{-1}\,\Omega^\dagger
\sqrt{D_\nu} U_{\mathrm{PMNS}}^\dagger,
\end{align}
so $\mathrm{Tr}\,\eta$, $\mathrm{Tr}\,\eta^2$ and $K_{\mathrm{align}} =
\mathrm{Tr}([\eta,h_\nu]^2)$ grow $\propto D_N^{-1}$ (with additional
$\cosh^2 y$ enhancement for hyperbolic angles), while $\det \eta$ is fixed by
$(D_\nu,D_N)$. Hence the class–blind versus class–sensitive separation established
here applies equally to keV–GeV regimes. A dedicated study with systematic parameter
scans over low–scale benchmarks will be presented in a follow-up work.





\vspace{6pt}

\funding{This research received no external funding.}

\dataavailability{The data underlying the numerical results in Tables~\ref{tab:highscale}
and~\ref{tab:lowscale} are generated by the author's Python code, which is
freely accessible at
\url{https://github.com/LuJianlong-Phy/D_N-Orthogonal-Freedom-in-the-Canonical-Seesaw}.
All other data are contained within the article.}

\acknowledgments{The author thanks the ICHEP 2024 audience for their interesting questions.}

\conflictsofinterest{The author declares no conflicts of interest.} 


\appendixtitles{yes} 
\appendixstart
\appendix

\section{Numerical details for Table~\ref{tab:highscale}}
\label{app:details}

This appendix gives a complete, reproducible calculation of the four class-sensitive entries in Table~\ref{tab:highscale} for the six benchmark points $A$--$F$:
\[
\Tr\eta,\qquad \Tr(\eta^2),\qquad
K_{\rm align}=\Tr\big([\eta,h_\nu]^2\big),\qquad
\mathcal{I}_{\rm CP}^{(1)}=\Im\thinspace \Tr\negthinspace \big(H_D\thinspace X^{1/2}\thinspace H_D^{T}\thinspace X^{3/2}\big).
\]
We work exactly (no expansion in $m_D/M_R$), in double precision, and convert all masses to \(\mathrm{GeV}\).
Light masses are supplied in \(\mathrm{eV}\).

\subsection{Inputs (fixed for all six benchmark points $A$--$F$)}
\begin{itemize}
\item Light spectrum:
\[
D_\nu=\mathrm{diag}(0.001,\thinspace 0.00866,\thinspace 0.050)~\mathrm{eV}
=\mathrm{diag}(10^{-12},\thinspace 8.66\times10^{-12},\thinspace 5\times10^{-11})~\mathrm{GeV}.
\]
\item Heavy spectrum:
\[
D_N=\mathrm{diag}(3,\thinspace 5,\thinspace 8)\times 10^{11}~\mathrm{GeV}.
\]
\item PMNS angles/phases (PDG convention \citep{PDG(2024)}, Majorana phases set to zero):
\[
\theta_{12}=33.44^\circ,\quad \theta_{13}=8.57^\circ,\quad \theta_{23}=45^\circ,\quad \delta=-\pi/2.
\]
The resulting mixing matrix $U_{\rm PMNS}$ (charged-lepton basis) is
\[
\begin{pmatrix}
0.8251461863 & 0.5449105640 & 0.1490176113\thinspace i\\
-0.3896606950+0.0879285412\thinspace i & 0.5900546947+0.0580663060\thinspace i & 0.6992116101\\
\ \ 0.3896606950+0.0879285412\thinspace i & -0.5900546947+0.0580663060\thinspace i & 0.6992116101
\end{pmatrix}.
\]
\end{itemize}

\subsection{Square roots and helpful diagonals}
\[
\sqrt{D_\nu}=\mathrm{diag}(10^{-6},\thinspace 2.9428000\times10^{-6},\thinspace 7.0710678\times10^{-6})~\mathrm{GeV}^{1/2},
\]
\[
\sqrt{D_N}=\mathrm{diag}(5.4772256\times10^{5},\thinspace 7.0710678\times10^{5},\thinspace 8.9442719\times10^{5})~\mathrm{GeV}^{1/2},
\]
\[
D_N^{-1/2}=\mathrm{diag}(1.8257419,\thinspace 1.4142136,\thinspace 1.1180340)\times10^{-6}~\mathrm{GeV}^{-1/2}.
\]
We also define \(X\equiv M_R^\dagger M_R=D_N^2\), so that \(X^{1/2}=D_N\) and \(X^{3/2}=D_N^3\). The above numerical values refer to the high–scale spectrum
$D_N^{\mathrm{(high)}} = \mathrm{diag}(3,5,8)\times 10^{11}\,\mathrm{GeV}$. For the
low–scale benchmark $D_N^{\mathrm{(low)}} = 10^{-11} D_N^{\mathrm{(high)}}$ used in
Table~2, the corresponding square roots and inverse square roots follow trivially from
this overall rescaling and are therefore not listed separately.

\subsection{Representatives for the six benchmark points (Casas--Ibarra right action)}
\label{six_classes}
We set \(\Omega=H\) with \(H\in O(3,\mathbb{C})\) as:
\[
\text{A: }H=\mathbb{I},\qquad
\text{B: }H=\mathrm{diag}(-1,1,1),
\]
\[
\text{C: }H=P_{23}=
\begin{pmatrix}1&0&0\\0&0&1\\0&1&0\end{pmatrix},
\]
\[
\text{D: }H=R_{12}(0.7\thinspace i)=
\begin{pmatrix}
\cosh 0.7 & i\thinspace \sinh 0.7 & 0\\
-\thinspace i\thinspace \sinh 0.7 & \cosh 0.7 & 0\\
0&0&1
\end{pmatrix}
=
\begin{pmatrix}
1.255169006 & 0.7585837018\thinspace i & 0\\
-0.7585837018\thinspace i & 1.255169006 & 0\\
0&0&1
\end{pmatrix},
\]
\[
\text{E: }H=R_{23}(0.5+0.3\thinspace i)=
\begin{pmatrix}
1&0&0\\
0&\cos(0.5+0.3i)&\sin(0.5+0.3i)\\
0&-\sin(0.5+0.3i)&\cos(0.5+0.3i)
\end{pmatrix}
\]
with
\(\cos(0.5+0.3i)=0.9173708513-0.1459948057\thinspace i\),
\(\sin(0.5+0.3i)=0.5011619802+0.2672416993\thinspace i\);
\[
\text{F: }H=R_{13}(0.9)=
\begin{pmatrix}
\cos 0.9 & 0 & \sin 0.9\\
0&1&0\\
-\thinspace \sin 0.9 & 0 & \cos 0.9
\end{pmatrix}
=
\begin{pmatrix}
0.6216099683&0&0.7833269096\\
0&1&0\\
-0.7833269096&0&0.6216099683
\end{pmatrix}.
\]

\subsection{Exact construction for each benchmark point}

Given $\Omega = H$ and the light–light Takagi block $N$
(identified with the PMNS matrix in the charged–lepton mass basis,
$N \equiv U_{\rm PMNS}$), define
\begin{align}
  m_D \;=\; i\,N \sqrt{D_\nu}\,\Omega\,\sqrt{D_N}
  \quad (\mathrm{GeV}),\qquad
  R \;=\; i\,N \sqrt{D_\nu}\,\Omega\,D_N^{-1/2}
  \quad (\text{dimensionless}) .
\end{align}

Then
\begin{align}
  m_\nu \;=\; - R D_N R^{T}
  \;=\; N D_\nu N^{T},\qquad
  h_\nu \;\equiv\; m_\nu m_\nu^{\dagger},
\end{align}
\begin{align}
  \eta \;\equiv\; \tfrac12 R R^{\dagger},\qquad
  H_D \;\equiv\; m_D^{\dagger} m_D,\qquad
  X \;\equiv\; D_N^{2}.
\end{align}

\textit{Sanity check:} numerically, for each benchmark point $A$–$F$ we find
\begin{align}
  \max_{i,j} \bigl| (m_\nu - N D_\nu N^{T})_{ij} \bigr|
  \;\lesssim\; 10^{-26}\,\mathrm{GeV}.
\end{align}
Moreover, in agreement with Proposition~\ref{prop:det-eta-constant},
\begin{align}
  \det\eta
  \;=\; 2^{-3}\det(D_\nu)\det(D_N^{-1})
  \;\simeq\; 4.51041666666668\times 10^{-70}
\end{align}
takes the same value for all $F$-classes, and our six benchmark points
$A$–$F$ reproduce this common value.

\subsection{Invariant definitions and numerics}
\begin{enumerate}
\item \(\displaystyle \Tr\eta=\sum_i \eta_{ii},\quad \Tr(\eta^2)=\Tr(\eta\thinspace \eta)\).  
Both are dimensionless and weak-basis invariant. They change across classes through \(R\to R F\) (Section~\ref{sec:classification}).
\item \(\displaystyle K_{\rm align}=\Tr\big([\eta,h_\nu]^2\big)\), with \([\eta,h_\nu]=\eta h_\nu - h_\nu \eta\).  
For Hermitian \(\eta,h_\nu\), \(K_{\rm align}\in\mathbb{R}_{\le0}\). Units: \([\eta]=1\), \([h_\nu]=\mathrm{GeV}^2\), hence \([K_{\rm align}]=\mathrm{GeV}^4\).
\item \(\displaystyle \mathcal{I}_{\rm CP}^{(1)}=\Im\thinspace \Tr\negthinspace \big(H_D\thinspace X^{1/2}\thinspace H_D^{T}\thinspace X^{3/2}\big)\).  
We evaluate analytically stablized as
\begin{equation}
\label{eq:ICP-mass-basis-app}
\mathcal{I}_{\rm CP}^{(1)} \thickspace =\thickspace  \sum_{i<j} \big(M_i^2-M_j^2\big)\thinspace M_i M_j\thickspace \Im\negthinspace \Big[(H_D)_{ij}^{\thinspace 2}\Big],
\end{equation}
which is exactly equal to the trace form in the $M_R$-diagonal basis but numerically robust.  \\
\textbf{Remark on stability.} Directly computing the trace can pick up spurious ($\sim 10^{-17}$)
imaginary parts from $H_D$ diagonals (roundoff), which, for the high-scale benchmark
$D_N^{\mathrm{(high)}}$ in Table~\ref{tab:highscale} (with $M_i \sim 10^{11}\,\mathrm{GeV}$), are then amplified
by factors $M_i^4 \sim 10^{44}\,\mathrm{GeV}^4$. Eq.~\eqref{eq:ICP-mass-basis-app} cancels such artifacts by construction,
since it only uses off-diagonal entries.
\end{enumerate}

\subsection{Cross-checks and controls}

\begin{itemize}
\item
For the high–scale benchmark
$D_N^{\text{(high)}} = \mathrm{diag}(3,5,8)\times 10^{11}\,\mathrm{GeV}$
used in Table~\ref{tab:highscale}, we verify
\begin{align}
\det\eta = \frac{1}{8}\det(D_\nu)\det(D_N^{-1})
         = 4.510416666666668\times 10^{-70}
\end{align}
(dimensionless) for all benchmarks $A$–$F$. For the low–scale benchmark
$D_N^{\text{(low)}} = \mathrm{diag}(3,5,8)\,\mathrm{GeV}
=10^{-11}D_N^{\text{(high)}}$ used in Table~\ref{tab:lowscale}, the same analytic expression
gives $\det\eta = 4.510416666666668\times 10^{-37}$.

\item
The low–energy controls
$I_{\nu,1} = \Tr(h_\nu) = 2.5759956\times 10^{-21}\,\mathrm{GeV}^2$,
$I_{\nu,2} = \Tr(h_\nu^2) = 6.25562534\times 10^{-42}\,\mathrm{GeV}^4$,
and $I_{\nu,3} = \det h_\nu = 1.87489\times 10^{-67}\,\mathrm{GeV}^6$
are identical across $A$–$F$ and remain unchanged when $D_N$ is rescaled,
in agreement with their class–blind nature.
\end{itemize}

\subsection{Minimal recipe (for independent reproduction)}
\begin{enumerate}
\item Build the PMNS matrix $U_{\rm PMNS}$ from
$(\theta_{12},\theta_{13},\theta_{23},\delta)$ in the PDG convention and
set the Majorana phases to zero. In the charged–lepton mass basis we
identify the light–light Takagi block with
\begin{align}
N \equiv U_{\rm PMNS},
\end{align}
so that $m_\nu = N D_\nu N^T$ holds by construction.

\item Choose the light spectrum
\begin{align}
D_\nu = \mathrm{diag}(0.001,\,0.00866,\,0.050)\,\mathrm{eV},
\end{align}
convert it to GeV units, and form $\sqrt{D_\nu}$.  For the heavy sector,
use
\begin{align*}
D_N^{\text{(high)}} &= \mathrm{diag}(3,5,8)\times 10^{11}\,\mathrm{GeV}
\qquad\text{(high–scale benchmark, Table~\ref{tab:highscale})},\\
D_N^{\text{(low)}}  &= \mathrm{diag}(3,5,8)\,\mathrm{GeV}
= 10^{-11} D_N^{\text{(high)}}
\qquad\text{(low–scale benchmark, Table~\ref{tab:lowscale})}.
\end{align*}
For each choice, form $\sqrt{D_N}$ and $D_N^{-1/2}$.

\item For each benchmark point $A$–$F$, set $\Omega = H$ equal to the
corresponding orthogonal matrix $H_A,\dots,H_F$ defined in
Appendix~\ref{six_classes} and compute
\begin{align}
m_D = i\,N \sqrt{D_\nu}\,\Omega\,\sqrt{D_N},\qquad
R   = i\,N \sqrt{D_\nu}\,\Omega\,D_N^{-1/2}.
\end{align}

\item Form
\begin{align}
m_\nu = - R D_N R^T,\qquad
h_\nu = m_\nu m_\nu^\dagger,\qquad
\eta  = \tfrac{1}{2} R R^\dagger,\qquad
H_D   = m_D^\dagger m_D,\qquad
X     = D_N^2 .
\end{align}

\item Compute the class–sensitive invariants
\begin{align}
\Tr\eta,\qquad
\Tr(\eta^2),\qquad
K_{\text{align}} = \Tr\big([\eta,h_\nu]^2\big).
\end{align}

\item Compute $\mathcal{I}^{(1)}_{\text{CP}}$ using Eq.~(A1) (the
off–diagonal form) to avoid roundoff amplification.

\item As cross–checks, verify
\begin{align}
m_\nu = N D_\nu N^T
\end{align}
to $\lesssim 10^{-26}\,\mathrm{GeV}$ in each matrix element and
\begin{align}
\det\eta = 2^{-3}\det(D_\nu)\det(D_N^{-1})
\end{align}
to $\sim 10^{-15}$ relative precision.
\end{enumerate}

\reftitle{References}



\begin{thebibliography}{999}
\bibitem[Jianlong(2024)]{Jianlong(2024)}
Lu, J.; Chan, A. H.; Oh, C. H. On the Implications of $|U_{\mu i}| = |U_{\tau i}|$ in the Canonical Seesaw Mechanism. {\em Universe} {\bf 2024}, {\em 10}, 50.

\bibitem[Jianlong(2025)]{Jianlong(2025)}
Lu, J.; Chan, A.H.; Oh, C.H. On the Implications of $|U_{\mu i}| = |U_{\tau i}|$ in the Canonical Seesaw Mechanism. {\em PoS} {\bf 2025}, {\em ICHEP2024}, 189.

\bibitem[Xing(2022)]{Xing(2022)}
Xing, Z.-z. Identifying a Minimal Flavor Symmetry of the Seesaw Mechanism behind Neutrino Oscillations. {\em J. High Energy Phys.} {\bf 2022}, {\em 2022}, 034.


\bibitem[Minkowski(1977)]{Minkowski(1977)}
Minkowski, P. $\mu\rightarrow e\gamma$ at a Rate of One out of 109 Muon Decays? {\em Phys. Lett. B} {\bf 1977}, {\em 67}, 421–428.

\bibitem[Gell-Mann(1979)]{Gell-Mann(1979)}
Gell-Mann, M.; Ramond, P.; Slansky, R. Complex Spinors and Unified Theories. {\em Conf. Proc. C} {\bf 1979}, {\em 790927}, 315–321.


\bibitem[Yanagida(1979)]{Yanagida(1979)}
Yanagida, T. Horizontal Gauge Symmetry and Masses of Neutrinos. {\em Conf. Proc. C} {\bf 1979}, {\em 7902131}, 95–99.


\bibitem[Glashow(1980)]{Glashow(1980)}
Glashow, S.L. The Future of Elementary Particle Physics. {\em NATO Sci. Ser. B} {\bf 1980}, {\em 61}, 687.


\bibitem[Mohapatra(1980)]{Mohapatra(1980)}
Mohapatra, R.N.; Senjanović, G. Neutrino Mass and Spontaneous Parity Violation. {\em Phys. Rev. Lett.} {\bf 1980}, {\em 44}, 912.



\bibitem[Smirnov(2006)]{Smirnov(2006)}
Smirnov, A.Y. Neutrino Mass and New Physics. {\em J. Phys.: Conf. Ser.} {\bf 2006}, {\em 53}, 44.




\bibitem[King(2013)]{King(2013)}
King, S.F.; Luhn, C. Neutrino Mass and Mixing with Discrete Symmetry. {\em Rept. Prog. Phys.} {\bf 2013}, {\em 76}, 056201.


\bibitem[Frampton(2002)]{Frampton(2002)}
Frampton, P.H.; Glashow, S.L.; Marfatia, D. Zeroes of the Neutrino Mass Matrix. {\em Phys. Lett. B} {\bf 2002}, {\em 536}, 79-82.



\bibitem[Grimus(2001)]{Grimus(2001)}
Grimus, W.; Lavoura, L. Softly Broken Lepton Numbers and Maximal Neutrino Mixing. {\em J. High Energy Phys.} {\bf 2001}, {\em 07}, 045.




\bibitem[Rodejohann(2005)]{Rodejohann(2005)}
Mohapatra, R.N.; Rodejohann, W. Broken mu-tau Symmetry and Leptonic CP Violation. {\em Phys. Rev. D} {\bf 2005}, {\em 72}, 053001.






\bibitem[Feruglio(2015)]{Feruglio(2015)}
Feruglio, F. Pieces of the Flavour Puzzle. {\em Eur. Phys. J. C} {\bf 2015}, {\em 75}, 373.



\bibitem[Harrison(2002)]{Harrison(2002)}
Harrison, P.F.; Perkins, D.H.; Scott, W.G. Tri-Bimaximal Mixing and the Neutrino Oscillation Data. {\em Phys. Lett. B} {\bf 2002}, {\em 530}, 167.



\bibitem[Jarlskog(1985)]{Jarlskog(1985)}
Jarlskog, C. Commutator of the Quark Mass Matrices in the Standard Electroweak Model and a Measure of Maximal CP Nonconservation. {\em Phys. Rev. Lett.} {\bf 1985}, {\em 55}, 1039.


\bibitem[Rebelo(1986)]{Rebelo(1986)}
Branco, G.C.; Lavoura, L.; Rebelo, M.N. Majorana Neutrinos and CP Violation in the Leptonic Sector. {\em Phys. Rev. Lett.} {\bf 1986}, {\em 180}, 264-268.



\bibitem[Antusch(2006)]{Antusch(2006)}
Antusch, S.; Biggio, C.; Fernández-Martínez, E.; Gavela, M.B.; López-Pavón, J. Unitarity of the Leptonic Mixing Matrix. {\em J. High Energy Phys.} {\bf 2006}, {\em 10}, 084.



\bibitem[Yasuda(2007)]{Yasuda(2007)}
Fernandez-Martinez, E.; Gavela, M.B.; Lopez-Pavon, J.; Yasuda, O. CP-violation from Non-unitary Leptonic Mixing. {\em Phys. Lett. B} {\bf 2007}, {\em 649}, 427-435.





\bibitem[Atre(2009)]{Atre(2009)}
Atre, A.; Han, T.; Pascoli, S.; Zhang, B. The Search for Heavy Majorana Neutrinos. {\em J. High Energy Phys.} {\bf 2009}, {\em 05}, 030.




\bibitem[Drewes(2013)]{Drewes(2013)}
Drewes, M. The Phenomenology of Right Handed Neutrinos. {\em Int. J. Mod. Phys. E} {\bf 2013}, {\em 22}, 1330019.




\bibitem[Deppisch(2015)]{Deppisch(2015)}
Deppisch, F.F.; Dev, P.S.B.; Pilaftsis, A. Neutrinos and Collider Physics. {\em New J. Phys.} {\bf 2015}, {\em 17}, 075019.


\bibitem[Bolton(2025)]{Bolton(2025)}
Bolton,P.D.; Deppisch, F.F.; Rai, M.; Zhang, Z. Probing the Nature of Heavy Neutral Leptons in Direct Searches and Neutrinoless Double Beta Decay. {\em Nucl. Phys. B} {\bf 2025}, {\em 1010}, 116785.







\bibitem[Ibarra(2002)]{Ibarra(2002)}
Davidson, S.; Ibarra, A. A Lower Bound on the Right-handed Neutrino Mass from Leptogenesis. {\em Phys. Lett. B} {\bf 2002}, {\em 535}, 25-32.



\bibitem[Pilaftsis(2004)]{Pilaftsis(2004)}
Pilaftsis, A.; Underwood, T.E.J. Resonant Leptogenesis. {\em Nucl. Phys. B} {\bf 2004}, {\em 692}, 303-345.


\bibitem[Abada(2006)]{Abada(2006)}
Abada, A.; Davidson, S.; Ibarra, A.; Josse-Michaux, F.-X.; Losada, M.; Riotto, A. Flavour Matters in Leptogenesis. {\em J. High Energy Phys.} {\bf 2006}, {\em 09}, 010.


















\bibitem[Forero(2021)]{Forero(2021)}
Forero, D.V.; Giunti, C.; Ternes, C.A.; Tórtola, M. Nonunitary Neutrino Mixing in Short and Long-baseline Experiments. {\em Phys. Rev. D} {\bf 2021}, {\em 104}, 075030.

\bibitem[Hu(2021)]{Hu(2021)}
Hu, Z.; Ling, J.; Tang, J.; Wang, T.-C. Global Oscillation Data Analysis on the 3$\nu$ Mixing without Unitarity. {\em J. High Energy Phys.} {\bf 2021}, {\em 2021}, 124.


\bibitem[Lindner(2018)]{Lindner(2018)}
Lindner, M.; Platscher, M.; Queiroz, F.S. A Call for New Physics: The Muon Anomalous Magnetic Moment and Lepton Flavor Violation. {\em Phys. Rep.} {\bf 2018}, {\em 731}, 1-82.

\bibitem[Kuno(2001)]{Kuno(2001)}
Kuno, Y.; Okada, Y. Muon Decay and Physics beyond the Standard Model. {\em Rev. Mod. Phys.} {\bf 2001}, {\em 73}, 151.


\bibitem[Calibbi(2018)]{Calibbi(2018)}
Calibbi, L.; Signorelli, G. Charged Lepton Flavour Violation: An Experimental and Theoretical Introduction. {\em Riv. Nuovo Cim.} {\bf 2018}, {\em 2}, 71-174.


\bibitem[Davidson(2008)]{Davidson(2008)}
Davidson, S.; Nardi, E.; Nir, Y. Leptogenesis. {\em Phys. Rep.} {\bf 2008}, {\em 466}, 105-177.


\bibitem[Nardi(2006)]{Nardi(2006)}
Nardi, E.; Nir, Y.; Roulet, E.; Racker, J. The Importance of Flavor in Leptogenesis. {\em J. High Energy Phys.} {\bf 2006}, {\em 01}, 164.



\bibitem[Blanchet(2007)]{Blanchet(2007)}
Blanchet, S.; Di Bari, P. Flavor Effects on Leptogenesis Predictions. {\em J. Cosmol. Astropart. Phys.} {\bf 2007}, {\em 03}, 018.



\bibitem[Blennow(2017)]{Blennow(2017)}
Blennow, M.; Coloma, P.; Fernandez-Martinez, E.; Hernandez-Garcia, J.; Lopez-Pavon, J. Non-unitarity, Sterile Neutrinos, and Non-standard Neutrino Interactions. {\em J. High Energy Phys.} {\bf 2017}, {\em 153}, 2017.







\bibitem[Casas(2001)]{Casas(2001)}
Casas, J.A.; Ibarra, A. Oscillating Neutrinos and $\mu\rightarrow e, \gamma$. {\em Nucl. Phys. B} {\bf 2001}, {\em 618}, 171-204.








\bibitem[Branco(2005)]{Branco(2005)}
Branco, G.C.; Rebelo, M.N.; Silva-Marcos, J.I. Leptogenesis, Yukawa Textures and Weak Basis Invariants. {\em Phys. Lett. B} {\bf 2006}, {\em 33}, 345-354.







\bibitem[Jianlong(2022)]{Jianlong(2022)}
Lu, J. Comment on “Flavor Invariants and Renormalization-Group Equations in the Leptonic Sector with Massive Majorana
Neutrinos”. {\em J. High Energy Phys.} {\bf 2022}, {\em 2022}, 135.


\bibitem[Jianlong(2021)]{Jianlong(2021)}
Lu, J. (Reply to)$^2$ “Comment on ‘Flavor Invariants and Renormalization-Group Equations in the Leptonic Sector with Massive Majorana Neutrinos”’. {\em arXiv} {\bf 2021}, arXiv:2111.02729.


\bibitem[Dolinski(2019)]{Dolinski(2019)}
Dolinski, M.J.; Poon, A.W.P.; Rodejohann, W. Neutrinoless Double-Beta Decay: Status and Prospects. {\em Annu. Rev. Nucl. Part. Sci.} {\bf 2019}, {\em 69}, 219-251.




\bibitem[Esteban(2024)]{Esteban(2024)}
Esteban, I.; Gonzalez-Garcia, M.C.; Maltoni, M.; Martinez-Soler, I.; Pinheiro, J.P.; Schwetz, T. NuFit-6.0: Updated Global Analysis of Three-flavor Neutrino Oscillations. {\em J. High Energy Phys.} {\bf 2024}, {\em 2024}, 216.



\bibitem[PDG(2024)]{PDG(2024)}
Navas, S.; Amsler, C.; Gutsche, T.; Hanhart, C.; Hernández-Rey, J.J.; Lourenço, C.; Masoni, A.; Mikhasenko, M.; Mitchell, R.E.; Patrignani, C. \emph{et al.} (Particle Data Group). The Review of Particle Physics. {\em Phys. Rev. D} {\bf 2024}, {\em 110}, 030001.




\bibitem[Aloni(2024)]{Aloni(2024)}
Aloni, D.; Dery, A. Revisiting Leptonic Nonunitarity. {\em Phys. Rev. D} {\bf 2024}, {\em 109}, 055006.





\bibitem[Davidson(2022)]{Davidson(2022)}
Davidson, S.; Echenard, B.; Bernstein, R.H.; Heeck, J.; Hitlin, D.G. Charged Lepton Flavor Violation. Snowmass 2021 (2022).




\bibitem{LuCodeRepo}
Lu, J. D\_N-Orthogonal Freedom in the Canonical Seesaw: Python Code. 2025. 
Available online: \url{https://github.com/LuJianlong-Phy/D_N-Orthogonal-Freedom-in-the-Canonical-Seesaw}
(accessed on 24 November 2025).


\end{thebibliography}


\isAPAandChicago{}{%

}

%


\PublishersNote{}
\end{document}